\documentclass[preprint,amsmath,amssymb]{revtex4}
\usepackage{graphicx}
\usepackage{dcolumn}
\usepackage{bm}
\begin{document}
\title{Dark fluid or cosmological constant : Why there are different de Sitter-type spacetimes}
\author{M.~Nouri-Zonoz  \footnote{nouri@ut.ac.ir, corresponding author}, J.~Koohbor,\footnote{jkoohbor@ut.ac.ir}  and H.~Ramezani-Aval  \footnote{hramezania@ut.ac.ir}} 
\affiliation {Department of Physics, University of Tehran, North Karegar Avenue, Tehran 14395-547, Iran.}
\begin{abstract}
Many different forms of the de Sitter metric in different coordinate systems are used in the general relativity literature. 
Two of them are the most common, the static form and the cosmological (exponentially expanding) form. The staticity 
and non-stationarity of these two different forms are traced back to the noncomoving and comoving nature of the 
corresponding coordinate systems. In this paper using the quasi-Maxwell form of the Einstein field equations and a definition of static spacetimes based upon them,  we look at these two different forms of the same solution from a new perspective which classifies them as a special case in a 
general one-parameter family of solutions. Specifically it is proved that, 
{\it irrespective of the spacetime symmetry, a one-element perfect fluid in any frame noncomoving with the fluid could be the source of a static spacetime, only if its equation of state is that of a dark fluid namely $p = -\rho = const.$}. These static solutions, which include the well-known de Sitter spacetime, are called de Sitter-type spacetimes. To exemplify we consider static axially and cylindrically symmetric de Sitter-type spacetimes and their dynamic (cosmological) versions. It is shown how despite the seemingly natural expectation based on the presence of $\Lambda$ as their only parameter, the nonspherical expansions of these genuinely different solutions should be expected indeed. To the best of our knowledge the dynamic version of the cylindrically symmetric de Sitter-type spacetime is introduced here for the first time. 
Finally it is noted that the identification of the geometric term $\Lambda g_{ij}$ with a perfect fluid with equation of state $p = -\rho = const.$, although mathematically consistent, obscures the crucial role of the (dark) fluid's velocity in defining a preferred (comoving) coordinate system in de Sitter-type spacetimes.
\end{abstract}
\maketitle
\section{Introduction and motivation}
In 1917, after Einstein's static universe \cite{Ein}, the de Sitter solution \cite{des} was the second cosmological model incorporating a new constant,
the so-called cosmological constant $\Lambda$ into the original Einstein field equations (EFEs). The main features of the two solutions were summed up in  Eddington's famous quote characterizing the Einstein static universe as {\it matter without motion} and that of de Sitter world as {\it motion  without matter}. Born in the same year, the two models had different fates. Einstein's static Universe turned to a case for pathological studies after Hubble's discovery of the expansion of the Universe, while the inflationary scenario and the recent discovery of the accelerated expansion of the Universe in the present epoch, resurrected interest in the cosmological constant ($\Lambda$). Specifically in the latter case $\Lambda$ is taken as the main candidate driving the repulsive gravity accounting for the unexpected observation. This in turn has led to a wide interest in  studying de Sitter and de Sitter-type spacetimes more carefully by putting their characteristics under detailed scrutiny. The nature of the cosmological term in modified Einstein field equations (MEFEs), either a universal (geometrical) constant or an exotic perfect fluid with an equation of state (EOS) $p= -\rho$ (the so-called dark energy) remains a fundamental question yet to be answered. The special case of a perfect fluid with $p= -\rho = const.$, is formally taken to be equivalent to a cosmological constant and this formal equivalency is one of the main subjects in the present study.
Historically de Sitter first introduced his solution in its static form
\begin{equation}\label{desit1}
ds^2 = (1-\frac{\Lambda r^2}{3}) c^2 dt^2 - (1-\frac{\Lambda r^2}{3})^{-1} dr^2 - r^2 d\Omega^2 
\end{equation}
but later it was realized that this solution is not written in a {\it comoving  synchronous} coordinate system (CSCS) \footnote{In a synchronous coordinate system metric components are such that $g_{00} = 1$ and $g_{0\alpha} = 0$, while in the comoving coordinate system the contravariant 4-velocity is given by $u^a = (1,0,0,0)$. Therefore in a CSCS we also have $u_a = (1,0,0,0)$. It can also be shown that the gravitational field cannot be stationary in a synchronous coordinate system \cite{Landau}.}. When it is transformed to such a coordinate system by the following transformations,
\begin{gather}
T = t + \frac{1}{2\sqrt{\frac{\Lambda}{3}}}\ln(1-\frac{\Lambda}{3} r^2) \label{trans11} \\ 
R= \frac{r}{\sqrt{1-\frac{\Lambda}{3} r^2}} e^{-\sqrt{\frac{\Lambda}{3}}t} \label{trans12}
\end{gather}
it is found to be the following exponentially expanding cosmological solution of EFEs,
\begin{equation}\label{desit2}
ds^2 = c^2 dT^2 - e^{2\sqrt{\frac{\Lambda}{3}}T} \left( dR^2 + R^2 d\Omega^2 \right),
\end{equation}
which is the flat FLRW model with an exponential expansion factor \footnote{In other words, using the geodesic equation, in the first form $\Gamma^i_{00} \neq 0$, while in the second form it is identically zero \cite{Landau}.}. It is noted that the time transformation \eqref{trans11} is nothing but the well-known Gullstrand-Painlev\'{e} transformation (coordinates) \cite{GP-1} which takes the static de Sitter metric to the following form,
\begin{equation}\label{desit21}
ds^2 = c^2 dT^2 - (dr -\sqrt{\frac{\Lambda }{3}}rc dT)^2 -r^2 d\Omega^2,
\end{equation}
in which the constant time hypersurfaces are flat and the fiducial observers (FIDOs) \cite{Thor} see the freely falling observers (FFOs) move radially outward at velocity $v_{esc} = \sqrt{\frac{\Lambda }{3}}rc$ \footnote{Actually, compared to the  Schwarzschild spacetime in the same coordinate system \cite{GP-3}, in the de Sitter case the freely falling observers should more appropriately be called freely escaping observers (FEOs), since they start with zero velocity at the center of the coordinates $r=0$ and reach the velocity of light at the de Sitter horizon $r=\sqrt{\frac{3}{\Lambda}}$.}. 
After the observation by Lanczos that a four- ($d$)-dimensional de Sitter space is a hyperboloid embedded in five- ($d+1$)-dimensional Minkowski spacetime \cite{Lanc}, different coordinate systems covering different patches on the hyperboloid (including \eqref{desit1}, \eqref{desit2} and \eqref{desit21}), were employed to realize this fact, each having its own merits and limitations \cite{Hawk,Spard}.
In this paper, our main objective is to  show that,
{\it irrespective of the spacetime symmetry, a one-element perfect fluid in a noncomoving frame could be the source of a static spacetime, only if its EOS is given by $p=-\rho = const.$}.\\
In this way, de Sitter-type solutions in their static forms are characterized as the only static, (one-element) perfect fluid solutions of EFEs in noncomoving frames. To be specific, by de Sitter-type spacetimes we mean those static solutions of the generalized vacuum EFEs $R_{ab} = \Lambda g_{ab}$ $(\Lambda > 0)$, the so-called (static) Einstein spaces, which reduce to the flat spacetime in the limit $\Lambda \rightarrow 0$. 
Further restriction to a special symmetry will lead to the static form of the corresponding de Sitter-type spacetime.
To achieve this goal we will employ a formulation of spacetime decomposition into spatial and temporal sections called the {\it threading} formalism (or $1+3$ splitting), through which, among other things,  EFEs could be expressed in the so-called  {\it quasi-Maxwell form} in a broader context called {\it gravitoelectromagnetism} \cite{LBNZ}.\\
The plan of the paper is as follows. In Sec. II we will introduce the $1+3$-splitting formalism and derive the quasi-Maxwell form of the EFEs in the presence of a perfect fluid. In Sec. III we show that the condition of staticity of an stationary spacetime is the absence of its gravitomagnetic field. Using this fact, in Sect. IV we examine the uniqueness of de Sitter-type spacetimes as the only static spacetimes of a one-element perfect fluid source in a noncomoving frame. In Secs. V and VI we discuss static axially  and cylindrically symmetric de Sitter-type spacetimes and find their time-dependent versions in the CSCS which are the axial and cylindrical counterparts of \eqref{desit2}, respectively.\\
Following Landau and Lifshitz \cite{Landau}, our convention for indices is such that Latin indices run from 0 to 3 while the Greek ones run from 1 to 3. We also keep $c$ but set $G=1$.
\section{$1+3$-splitting and the Quasi-Maxwell form of the Einstein filed equations}
The main idea of any  splitting formalism in general relativity is the introduction of spatial and temporal sections of a  spacetime metric so that one could assign spatial distances and time intervals to nearby events. There are two well-known separation formalisms: $3+1$ or the {\it foliation} formalism \cite{MTW} and $1+3$ or the {\it threading} formalism. In the threading formulation of spacetime decomposition with which we are concerned, the spacetime metric is expressed in the following general form \cite{Landau};
\begin{equation}\label{metric1}
ds^2 = c^2 d\tau_{syn}^2 - dl^2 = g_{00}(dx^0 - A_\alpha dx^\alpha)^2-\gamma_{\alpha\beta} dx^\alpha dx^\beta, \;\;\; \alpha, \beta = 1,2,3
\end{equation}
where  $A_\alpha = -\dfrac{g_{0\alpha}}{g_{00}}$ and  $\gamma_{\alpha\beta}= -g_{\alpha\beta}+\dfrac{g_{0\alpha}g_{0\beta}}{g_{00}}$
is the spatial metric of a 3-space, $\Sigma_3$,  on which $dl^2$ gives the infinitesimal spatial distance between any two events \footnote{It should be noted that $\Sigma_3$ is a differentiable 3-manifold but not a hypersurface in the original spacetime manifold with metric (\ref{metric1}). Indeed it is called the quotient space $\frac{\cal M}{G_1}$ where $G_1$ is the one-dimensional group of motions generated by the timelike Killing vector field of the spacetime under consideration \cite{Steph} .}. Also $d\tau_{syn} = \frac{1}{c}\sqrt{g_{00}}(dx^0-A_\alpha dx^\alpha)$ gives the infinitesimal interval of the so-called {\it synchronized proper time} between any two events at spatial distance $dl$.
This is obviously different from the proper time $d\tau =\frac{1}{c}\sqrt{g_{00}}dx^0$ measuring the time interval between two events happening at the same point($dx^\alpha=0$). In static spacetime these two concepts coincide but in stationary spacetimes they differ due to the presence of the metric cross terms ($ g_{0\alpha}$) which result in a further nonzero coordinate-time separation (desynchronization) 
\begin{equation}\label{TD}
\Delta x^0 = -\frac{g_{0\alpha}}{g_{00}}dx^\alpha 
\end{equation}
between simultaneous events at nearby spatial points. To be able to synchronize clocks all over space, one should choose a coordinate system in which the cross terms $g_{0\alpha}$ vanish \cite{Landau}. This is obviously satisfied in a synchronous coordinate system in which, due to the  fact that $g_{00}=1$, the coordinate time coincides with the proper time measured at each point. The above coordinate time difference should be accounted for by the observers who for example are going to assign a 3-velocity to a test particle passing them by in a stationary spacetime (see Fig. 1). Indeed using this formulation, the 3-velocity of a particle in static and stationary spacetimes is given by $v^\alpha = \frac{dx^\alpha}{d\tau}$ and 
\begin{eqnarray}\label{11}
v^\alpha = \frac{dx^\alpha}{d\tau_{syn}} = \frac{c dx^\alpha}{\sqrt{g_{00}}(dx^0 - {A_\alpha} dx^\alpha)},
\end{eqnarray} 
respectively \cite{Landau}.
\begin{figure}
\begin{center}
\includegraphics[angle=0,scale=0.6]{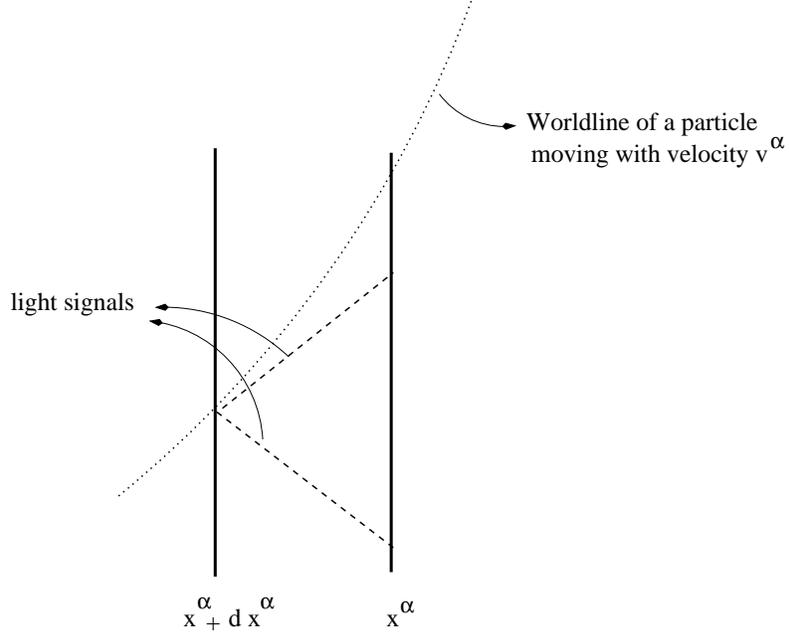}
\caption{A light signal sent and received back between two observers fixed at infinitesimally close points  $x^\alpha$ and $x^\alpha + dx^\alpha$ while a particle with 3-velocity $v^\alpha$ crosses their worldlines.}
\end{center}
\end{figure}
Also the components of the 4-velocity $u^i = \frac{dx^i}{ds} \;\;(i=0,1,2,3)$ of a test particle, in terms of the components of its 3-velocity are given by
\begin{eqnarray}\label{13}
u^0 =\frac{1}{\sqrt{g_{00}}\sqrt{1-v^2/c^2}} + \frac{A_\alpha v^\alpha}{\sqrt{1-v^2/c^2}}~~~;~~~u^\alpha = \frac{v^\alpha}{\sqrt{1-v^2/c^2}}.
\end{eqnarray}
In the $1+3$ decomposition of a stationary spacetime, one can define the so-called  {\it gravitoelectric} 
and {\it gravitomagnetic} fields, in terms of the derivatives of the metric components given by \eqref{metric1}, as follows \cite{LBNZ}
\begin{gather}
\textbf{E}_g=-\dfrac{\nabla h}{2h} \;\;\; , \;\;\;
\textbf{B}_g=\nabla \times \textbf{A} \label{B1}.
\end{gather}
in which $h \equiv g_{00}$. In terms of the above fields, EFEs for a one-element perfect fluid source could be written in the following {\it quasi-Maxwell} form \cite{LBNZ}
\begin{gather}
\nabla \times ~\textbf{E}_g=0, ~~~\nabla \cdot  \textbf{B}_g=0 \\
\nabla \cdot \textbf{E}_g= \frac{1}{2} h B^2_g+E^2_g - \frac{8\pi}{c^4}\left(\dfrac{p+\rho}{1-\frac{v^2}{c^2}}-\dfrac{\rho-p}{2}\right) \label{r00}\\  \label{r01}
\nabla \times  (\sqrt{h}\textbf{B}_g)=2 \textbf{E}_g \times (\sqrt{h}\textbf{B}_g)-\frac{16\pi}{c^4}\left(\dfrac{p+\rho}{1-\frac{v^2}{c^2}}\right) \frac{\textbf{v}}{c} \\
{^{(3)}}P^{\mu\nu}=-{E}_g^{\mu;\nu}+\frac{1}{2}h(B_g^\mu B_g^\nu - B_g^2 \gamma^{\mu\nu})+ {E}_g^\mu E_g^\nu+ \frac{8\pi}{c^4}\left(\dfrac{p+\rho}{c^2 - v^2}v^\mu v^\nu+\dfrac{\rho-p}{2}\gamma^{\mu\nu}\right), \label{3ricci}
\end{gather}
where ${^{(3)}}P^{\mu\nu} $ is the three-dimensional Ricci tensor of the 3-space $\Sigma_3$ and all the differential operations are defined in the same space \cite{Landau,LBNZ}. It should also be noted that in the above expressions, the 3-velocity $v^\mu$ of the perfect fluid elements is defined according to \eqref{11}.
\section{Staticity condition in terms of the absence of the gravitomagnetic field}
In this section, our main goal is to prove the following statement:\\
{\it A stationary spacetime is static if and only if its gravitomagnetic field vanishes, i.e  ${\bf B}_g =0$.}\\
By its  mathematical definition, a spacetime is  stationary if it possesses a timelike Killing vector field and if the same Killing vector field is also hypersurface orthogonal, then that spacetime would be a static one.
To prove the above  statement, we first note that the gravitoelectromagnetic 3-vector fields defined in the previous section could be elevated to the following covariant 4-vector fields  \cite{NP14}
\begin{gather}
{E_g}_a = -\frac{1}{2|\xi|} {F_g}_{ab} \frac{\xi^b}{{\vert\xi\vert}}\\
B_g^a = -\vert\xi\vert \xi^b \eta_b^{a m n} (\frac{\xi_n}{{\vert\xi\vert}^2})_{;m} = -\frac{1}{2 |\xi|^2}\eta^{a b n m} \frac{\xi_b}{\vert\xi\vert} {F_g}_{nm} \label{b}
\end{gather}
in which, in analogy with the definition of the electromagnetic field tensor in curved spacetimes, the gravitoelectromagnetic field tensor $F_g$ (or Papapetrou field  \cite{NP14}) is defined as (also refer to Sec. $18.1$ of Ref. \cite{Steph})
\begin{equation}
{F_g}_{ab} = - |\xi|^2 \eta_{n m  a b}\frac{\xi^n}{|\xi|} B_g^m+2(\xi_a {E_g}_b-\xi_b {E_g}_a )\label{f-deco}
\end{equation}
where $\eta_{n m  a b}= \sqrt{-g} \epsilon_{n m  a b}=\sqrt{h} \sqrt{\gamma}\epsilon_{n m  a b}$ is the four-dimensional Levi-Civita pseudotensor. 
Obviously from the above definitions we have  ${E_g}_a\xi^a=0$ and ${B_g}_a\xi^a=0$; i.e. they ensure that these 4-vectors have no components along the timelike Killing vector field. In other words in the coordinate system adapted to the Killing vector they reduce to 3-vectors, i.e. ${E_g}_0 \doteq 0 \doteq {B_g}_0$ as expected. 
As pointed out, in mathematical jargon, a spacetime is called static if it possesses a hypersurface orthogonal timelike Killing vector field. This is guaranteed if the Killing vector satisfies the relation,
\begin{equation}\label{twist}
\xi_{[a;b}\xi_{c]}=0.
\end{equation}
So to prove our statement we need to show that the vanishing of ${B}_g$ is equivalent to the above relation. To this end we rewrite Eq. \eqref{b} as follows;
\begin{gather}
B_g^a = -\frac{1}{\vert\xi\vert} \xi^b \eta_b^{a m n} {\xi_n}_{;m}  - {\vert\xi\vert}{\xi_b} {\xi_n} \eta^{b a m n} {\xi_n} (\frac{1}{{\vert\xi\vert}^2})_{;m} \nonumber\\
      = -\frac{1}{\vert\xi\vert} \omega^a 
\end{gather}
in which the second term in the first line vanishes  due to the antisymmetricity of $\eta^{b a m n}$  and $ \omega^a =  \xi^b \eta_b^{a m n} {\xi_n}_{;m} $ is the so-called twist of $\xi_a$ \cite{Geroch}. In other words the gravitomagnetic 4-vector is proportional to the twist of the timelike Killing vector which, by Eq. \eqref{twist}, measures the failure of the Killing vector to be hypersurface orthogonal \cite{Wald}, hence proving our statement. 
The above criterion helps one to find out whether a spacetime metric, in an apparently stationary form, is static or not. The simple prescription is: (i) find the gravitomagnetic potential $A^\alpha_g$ of the spacetime metric by writing it in the form \eqref{metric1} and (ii) calculate the corresponding gravitomagnetic field using \eqref{B1}; the spacetime is static if this quantity vanishes. Examples of the application of this criterion of staticity include the Schwarzschild and de Sitter spacetimes in the Gullstrand-Painlev\'{e} coordinates \cite{GP-3}, in which their gravitomagnetic potentials are shown to be curl-free.
\section{Static spacetimes in noncomoving frames}
Now that we have established the staticity condition in terms of the nonexistence of the gravitomagnetic field of the underlying stationary spacetime we are only one step away from what we mentioned as one of the main objectives of the present study. To get there we draw the reader's attention to an
interesting feature in the quasi-Maxwell form of the EFEs which is the simple fact that by Eq. (\ref{r01}),
\begin{equation}
\nabla \times  (\sqrt{h}\textbf{B}_g)=2 \textbf{E}_g \times (\sqrt{h}\textbf{B}_g)-\frac{16\pi}{c^4}\left(\dfrac{p+\rho}{1-\frac{v^2}{c^2}}\right) \frac{\textbf{v}}{c} \nonumber
\end{equation}
a static (i.e. $\textbf{B}_g = 0$) solution produced by a one-element perfect fluid source, {\it in general} has to be in the comoving frame ($ \textbf{v} = 0 $) with respect to the fluid particles. In other words in this case the fluid elements follow the timelike congruence defined by the timelike Killing vector field in the adapted coordinate system in which $\xi^a \doteq (1,0,0,0)$. As an example of this case one could mention the well-known static interior Schwarzschild solution which is obtained in the comoving (but not synchronous) coordinate system \footnote{In the case of the interior Schwarzschild solution which is the simplest model of a star interior, $\Gamma^i_{00} \neq 0$ and this is due to the fact that, for a noncollapsing star, the perfect fluid elements are not moving on timelike geodesics. This is also clear from the dependence of fluid's pressure on the radial coordinate \cite{Steph1} leading to a pressure gradient which in turn forbids the introduction of a comoving synchronous coordinate system \cite{Landau}.}. An obvious exception in the above feature is the case where the EOS of the perfect fluid is that of {\it dark energy}, i.e $p = -\rho = const.$ \footnote{It is noted that one in principle could also have $p(x^\alpha) = -\rho(x^\alpha)$, i.e the pressure (density) of the fluid  could be position-dependent, but for the transformed coordinate system to be a CSCS it is required that the pressure gradient should vanish \cite{Landau}. This also allows the perfect fluid to act formally as the cosmological constant.}, where now one could have static solutions even in the noncomoving frame. One may raise the question that in this case it is expected that by solving the quasi-Maxwell equations (which are equivalent to EFEs), the nonzero velocity of the perfect fluid should enter the spacetime metric, which is obviously not the case with static de Sitter(-type) spacetime(s). But a closer examination of the quasi-Maxwell equations reveals the very simple but important  feature that by setting $p = -\rho$, all the velocity-dependent components of the fluid's energy-momentum tensor disappear ($T^{ab} = (p+\rho)u^au^b - pg^{ab}$) and so the velocity of the perfect fluid is not expected to be present in the spacetime metric, as is the case with de Sitter(-type) spacetime(s). Now by transforming to the CSCS, the 3-velocity of the fluid is set to zero, but as shown in getting from \eqref{desit1} to \eqref{desit2}, this happens at the expense of the metric transforming to its dynamical form. This could be seen explicitly by noting that the application of the time transformation \eqref{trans11}, as shown previously, transforms the metric to its Gullstrand-Painlev\'{e} form in the coordinate system corresponding to the proper time of freely escaping observers along the {\it outgoing radial timelike geodesics} of \eqref{desit1} which also represent the trajectories of the fluid elements. This is so because this transformation leads to the coordinate system in which $u_a = (1,0,0,0)$ \cite{GP-3}, while the radial coordinate transformation \eqref{trans12}, takes the metric to its synchronous form in CSCS (i.e $g_{00} =1$ and  $g_{0\alpha} =0$) where  now $u^a = (1,0,0,0)$ \footnote{Note that, whatever the nature of this fluid, it is taken to be a timelike entity, i.e. its elements move on timelike trajectories (geodesics).}. On the other hand due to the vanishing of the pressure gradient for the perfect fluid with EOS $p = -\rho= const.$, a synchronous coordinate system could also be comoving, one in which the perfect fluid elements are at rest \cite{Landau}. In the next two sections it will be shown that this is the general procedure which leads to the dynamic versions of de Sitter-type spacetimes starting from their static counterparts and, by the same token, allows a physically consistent interpretation for their anisotropic expansions. \\
In summary, the above result is a remarkable one; it shows that the static de Sitter-type spacetimes are one of a kind. They are the only static solutions of the EFEs with a one-element perfect fluid and in a frame {\it noncomoving} with the fluid.
\section{axially and cylindrically symmetric  de Sitter-type spacetimes}
To reinforce the above interpretation of the dynamic de Sitter-type spacetimes, in this section we consider axially and cylindrically symmetric  de Sitter-type spacetimes in their static and dynamic forms.
\subsection{Nariai spacetime as an axially expanding Universe}
As the first  example of a de Sitter-type spacetime we consider the Nariai metric  which is given by the following line element in  {\it noncomoving} spherical coordinates \cite{Nariai,Steph},
\begin{equation}\label{desit3}
ds^2 = (1-{\Lambda r^2}) c^2 dt^2 - (1-{\Lambda r^2})^{-1} dr^2 - \frac{1}{\Lambda} (d\theta^2 +\sin^2\theta d\phi^2) \;\; (\Lambda > 0)
\end{equation}
which is a product space $dS_2 \times S^2$. In the CSCS, its dynamic version (called Berttoti-Kasner space by Rindler \cite{Rind}) is given by the following line element;
\begin{equation}\label{desit4}
ds^2 = c^2 dT^2 - e^{2\sqrt{\Lambda}T} dR^2 - \frac{1}{\Lambda} (d\theta^2 +\sin^2\theta d\phi^2).
\end{equation}
As discussed by Bonnor \cite{bonn} and Rindler \cite{Rind}, $T=constant$ hypersurfaces of the above metric are homogeneous 3-cylinders of constant radius $\frac{1}{\sqrt{\Lambda}}$ and consequently interpreted as a uni-directionally expanding  spacetime. To show how this interpretation is a natural outcome of transforming the static version to a CSCS, we start by rewriting the static metric \eqref{desit3} in cylindrical coordinates $(t,z,\rho,\phi)$  with the usual ranges, in either of the following alternative forms \cite{Rind},
\begin{equation}\label{desit3-1}
ds^2 = (1-{\Lambda z^2}) c^2 dt^2 - (1-{\Lambda z^2})^{-1} dz^2 - \frac{1}{\Lambda} (d\rho^2 +\sin^2\rho d\phi^2)
\end{equation}
\begin{equation}\label{desit3-2}
ds^2 = (1-{\Lambda z^2}) c^2 dt^2 - (1-{\Lambda z^2})^{-1} dz^2 - \frac{1}{(1+\frac{\Lambda}{4} {\bar{\rho}}^2)^2} ({d{\bar{\rho}}^2} +{\bar{\rho}}^2 d\phi^2)
\end{equation}
The second form is obtained by a stereographic projection of $S^2$ in \eqref{desit3} onto $R^2$ (in polar coordinates), leading to spacetime topology of $dS_2\times R^2$, which manifests its axially symmetric nature. This form also shows clearly that by our definition Nariai spacetime is a de Sitter-type spacetime, i.e reduces to flat spacetime as $\Lambda \rightarrow 0$. Now using the following transformations 
\begin{gather}
T = t + \frac{1}{2\sqrt{\Lambda}}\ln(1-\Lambda z^2) \label{trans2-1} \\ 
Z= \frac{z}{\sqrt{1-\Lambda z^2}} e^{-\sqrt{\Lambda}t}, \label{trans2-2}
\end{gather}
the static metric \eqref{desit3-2} could be written in the CSCS ($T, Z, \rho, \phi$) as follows,
\begin{equation}\label{desit3-3}
ds^2 = c^2 dT^2 - e^{2\sqrt{\Lambda}T} dZ^2 - \frac{1}{(1+\frac{\Lambda}{4} {\bar{\rho}}^2)^2} ({d{\bar{\rho}}^2} +{\bar{\rho}}^2 d\phi^2)
\end{equation}
As in the case of the de Sitter spacetime, the transformation to the CSCS was obtained in two steps. In the first step the Gullstrand-Painelev\'{e} transformation \eqref{trans2-1} is employed to transform to a coordinate system moving along the axial ($z$-directed) timelike geodesics of \eqref{desit3-2}, corresponding to the freely escaping observers who start with zero velocity at $z=0$. This is  followed by a synchronous transformation implemented by \eqref{trans2-2}. These transformations justify the  interpretation of the dynamic version of the spacetime  as an {\it axially} expanding universe. In studying this spacetime, Rindler after saying that:\\
{``\it ..a $\Lambda$ term in the field equations is tantamount to the energy tensor of an exotic but isotropic fluid.''} \\
poses the following question:\\ 
``{\it How can isotropic sources ``cause'' a one-directional field? Is this another example of an anti-Machian universe, i.e. one whose spacetime symmetries are incompatible with the symmetries of its source?}``\\
By the above arguments our answer to this question is clear. The spacetime symmetries are compatible with the symmetries of its both dark and nondark sources. In this case while there is no nondark source for the field, it possesses a dark source which is a perfect fluid with an EOS $p=-\rho$ and a unidirectional (bulk) motion. This motion defines a distinct CSCS in which the dynamic version of the metric is given by \eqref{desit3-3} and whose unidirectional expansion is naturally dictated by the dark fluid's velocity.
\subsection{Cylindrically symmetric de Sitter-type spacetime}
To exemplify the above interpretation further and as another example of a de Sitter-type spacetime we consider the following static solution to the MEFEs;
\begin{equation}\label{cyldes}
 ds^2={\cos^{4/3}\bigg(\frac{\sqrt{3\Lambda}}{2}\rho\bigg) (d t^2
 - d z^2) - d\rho^2 - \frac {4}{3\Lambda}
 \sin^2\bigg(\frac{\sqrt{3\Lambda}}{2}\rho\bigg)
 \cos^{-2/3}\bigg(\frac{\sqrt{3\Lambda}}{2}\rho\bigg) d\phi^2},
\end{equation}
which could be obtained from the spacetime metric of a cylindrical distribution of matter in the presence of the cosmological constant, by setting the 
linear mass density equal to zero \cite{Linet, Tian, Grif}. Again as in the previous example, the apparently natural expectation that this solution should be the usual de Sitter spacetime, perhaps in a different coordinate system, is not fulfilled \footnote{In Ref. \cite{Grif} the authors explain this expectation by saying that  {``\it{Interestingly, the ``no source'' limit $\sigma = 0$ is not the (anti-)de Sitter space, as one would naturally expect!.}``}} . Indeed it is yet another genuinely different de Sitter-type spacetime (it goes over to the flat spacetime as $\Lambda \rightarrow 0$) as one can see by calculating its scalar invariants including Kretschmann invariant given by,
\begin{equation}
K \equiv R^{abcd}R_{abcd}=\frac{4}{3} \frac{(2\cos^{4}\big(\frac{\sqrt{3\Lambda}}{2}\rho\big) + 1)}{
\cos^{4}\big(\frac{\sqrt{3\Lambda}}{2}\rho\big)}\Lambda^2 = \frac{8}{3} \Lambda^2 + \frac{4}{3} \Lambda^2 \cos^{-4}\big(\frac{\sqrt{3\Lambda}}{2}\rho\big),
\end{equation}
which is obviously different from the Kretschmann invariants $K= \frac{8}{3} \Lambda^2$ and $K= 8\Lambda^2$  of the de Sitter and Nariai solutions respectively.
This is a solution which could be obtained by applying cylindrical symmetry to the quasi-Maxwell form of the EFEs after setting $B_g = 0$ and taking $p = -\rho = const.$, again with the condition that at the limit $\Lambda \rightarrow 0$ it goes over to the Minkowski spacetime.
Interestingly enough, it is unnoticed in the exact solution literature \cite{Steph,Gri} that the above solution can be obtained as a special case ($\gamma = -1$) of static cylindrically symmetric perfect fluid solutions with  barotropic EOS $p=\gamma \rho$ in the following equivalent form,
\begin{gather}\label{cyldes-1}
ds^2=F^{2/3} (d t^2 - d z^2) - F^{-1} d \bar{\rho}^2 -  F^{-1/3}{\bar{\rho}^2  d\phi^2}\\
\end{gather}
with $F= 1-\frac{3}{4}\Lambda \bar{\rho}^2$, through the following transformation
\begin{equation}
\bar\rho = \frac{2}{\sqrt{3\Lambda}} \sin\big(\frac{\sqrt{3\Lambda}}{2}\rho\big),
\end{equation}
in which $0 < \bar{\rho} < \frac{2}{\sqrt{3\Lambda}}$ for  $0 < {\rho} < \infty$. The above solution shows how the 3-velocity of the perfect fluid, in spite of its absence in the metric components in the noncomoving frame, affects the solution by selecting a preferred direction. In the case of the usual de Sitter spacetime, the static form gives the solution in a noncomoving coordinate system where each observer finds itself at the center of the coordinate system from which the de Sitter horizon is measured, but in the CSCS the comoving observer discovers that the elements of the dark fluid (or for that matter spacetime points) are running away {\it radially} ({\it isotropically}). In the same way the above cylindrically symmetric spacetime is interpreted in terms of the noncomoving coordinate system in which the dark fluid has a velocity in a preferred direction, here along the radial cylindrical coordinate. Now the question arises as what would be the dynamic form of the above spacetime metric in the CSCS?. In other words what is the cylindrical counterpart of spacetime metrics \eqref{desit2} and \eqref{desit3-3}?. To find the answer one could follow the same procedure employed in finding the time-dependent metrics \eqref{desit2} and \eqref{desit3-3} in previous sections and arrive at the following cylindrically symmetric dynamic spacetime in CSCS,
\begin{equation}\label{cyldes-2}
ds^2= dT^2 -  F^{-1/3}(\tilde{\rho}, T) B(\tilde{\rho}, T) (d \tilde{\rho}^2 + {\tilde{\rho}^2  d\phi^2}) - F^{2/3}(\tilde{\rho}, T) d z^2
\end{equation}
through the  transformations
\begin{gather}\label{TR1}
dt = F^{-2/3}dT + {A B}{F^{-1}} d \tilde\rho \\
d\bar\rho = A(\tilde{\rho}, T) dT  + B(\tilde{\rho}, T) d \tilde\rho,
\end{gather}
where the functions $A$ and $B$ are given, respectively, by
\begin{gather}\label{TR2}
A = (F^{1/3} - F)^{1/2}\\
\int \frac{d B}{\sqrt{(1-\frac{3}{4}\Lambda {\tilde\rho^2} B^2 )^{1/3} - (1-\frac{3}{4}\Lambda {\tilde\rho^2} B^2 )}}= \frac{T}{\tilde\rho}.
\end{gather}
The above spacetime is a cylindrically expanding universe where the expansion factors in the polar plane and along the z direction are different, indicating a nonisotropic expansion. To the best of our knowledge the above dynamic version of the cylindrically symmetric static de Sitter-type spacetime is introduced here for the first time.
As a final remark it should be noted that as in the case of the usual de Sitter spacetime, it is straightforward to show that all the arguments made here could be repeated for a negative cosmological constant. Indeed a negative $\Lambda$ {\it static} solution (anti-de Sitter-type spacetime) of the above nature is discussed in Ref. \cite{Bonnor}.
\section{Conclusion}
In the present study after introducing the $1+3$ or threading formulation of spacetime decomposition 
it was shown that a stationary spacetime is static if and only if its gravitomagnetic field vanishes.
Applying this definition of staticity to the quasi-Maxwell form of the EFEs, we obtained the main objective of the paper that 
{\it a one-element perfect fluid in a noncomoving frame could be the source of a static spacetime, only if its EOS is that of a dark fluid namely $p = -\rho = const.$}.\\
The above assertion shows that de Sitter-type spacetimes are unique solutions and further, implies why there should be different de Sitter-type spacetimes.
To exemplify, comparing the three  solutions \eqref{desit1}, \eqref{desit3-2} and \eqref{cyldes}, it was shown that they are genuinely different solutions of vacuum MEFE which, by setting $\Lambda = 0$, all reduce to the Minkowski spacetime in different coordinate systems. One may now ask why, in the absence of matter, there is a unique flat spacetime solution of EFEs but genuinely different de Sitter-type solutions of MEFEs? Our answer to this question focuses on the fact that there is a hidden parameter which distinguishes between different de Sitter-type solutions and that is the velocity of the perfect fluid with EOS $p = -\rho = const.$  which formally plays the role of the cosmological constant in these solutions. Indeed the expectation that by setting the parameter characterizing the mass distribution in the Linet-Tian solution equal to zero, it should reduce to the usual de Sitter spacetime, arises from confining the symmetry of the spacetime only to its nondark sources, whereas one should take into account the symmetry of its dark sources which is  a perfect fluid with EOS $p = -\rho = const.$. To further clarify this point  consider Shcwartzschild and Levi-Civita solutions which are one-parameter static spherically and cylindrically symmetric solutions of vacuum EFEs respectively. In both  solutions setting the mass/mass per unit length parameter equal to zero we end up with the flat spacetime, whereas in the cases of the Schwartzschild-de Sitter and Linet-Tian solutions setting the mass/mass per unit length parameter equal to zero  we end up with two different solutions with apparently the same geometrical parameter which is the cosmological constant/term.
Borrowing Eddington's language our assertions could be summed up as follows: Einstein's static universe is a universe in a coordinate system with ``{\it comoving matter}'', time-(in)dependent de Sitter-type spacetimes are universes in a coordinate system with a ``{\it (non)comoving dark fluid}'' and Minkowski spacetime is the ``{\it no matter, no dark fluid}'' universe. Obviously it is this ``{\it no nothing}'' feature of the flat spacetime which makes it a unique solution of EFEs in the absence of matter and,  by the same token, it is the nonzero velocity of a dark fluid which produces different de Sitter-type spacetimes. The crux of the matter is that as soon as you model the cosmological constant by a perfect fluid, you are in principle assigning three different quantities to it: density, pressure and a 4-velocity. On the other hand, characterizing this fluid by the equation of state $p = -\rho = const.$  not only makes the first two quantities dependent but also, at the same time, hides the role of its 4-velocity both in the fluid's energy-momentum tensor and in the corresponding cosmological solution. The above studies show in a remarkable way, how the role of this exotic fluid's 4-velocity manifests itself in the anisotropic expansion of the dynamic versions of the (nonspherical) de Sitter-type spacetimes. Following this line, as an important by-product of this study, we introduced the dynamic  version of the cylindrically symmetric de Sitter-type spacetime. \\
In brief the above arguments show that the identification of the geometric (cosmological constant) term $\Lambda g_{ij}$, with a perfect fluid with the EOS $p = -\rho = const.$, is somewhat misleading due to the fact that the role of the fluid's velocity in characterizing the nature of the corresponding de Sitter-type spacetime is eclipsed in this identification. In other words for a consistent interpretation of de Sitter-type spacetimes in their dynamic forms, one has to employ the perfect fluid model. 
\section *{Acknowledgments}
The authors would like to thank University of Tehran for supporting this project under the grants provided by the research council. M. N-Z also thanks M. Noorbala, A. Parvizi and M. M. Sheikh-Jabbari for useful discussions.



\end{document}